# The solution of many-electron problems arising when radiation pulses interact with closed- or open-shell multielectron atomic and molecular states          September 3 2017

## Cleanthes A. Nicolaides

*Theoretical and Physical Chemistry Institute, National Hellenic Research Foundation, Athens 11635, Greece     e-mail: caan@eie.gr*

## Prologue

This paper was submitted to J Phys B on July 31, 2017, as a brief review and commentary on aspects of our work on the title theme, in response to a 'call for papers' for publication in a special issue on '*Correlations and light-matter interactions*'. It was rejected, the argument being that, '*The author does not describe a specific question that is solved in the current work*'.

It is a fact that the paper does not solve a specific new problem. Instead, it gleans from published work by the author and his colleagues on the title theme. The information which is summarized here is timely and useful, in the context of theoretical/computational and experimental research on various types of **time-independent** or **time-dependent many-electron problems** arising from the interaction of radiation with atomic and molecular states.

**Abstract:** When a pulse of radiation interacts with an atomic or molecular state, the observable quantities contain the information from the *self-consistent fields* and from the *electron correlations* in the wavefunctions of the physically and computationally relevant states. Depending on the nature of the interaction, e.g., on the duration and intensity of the pulse, the formal description and the ab initio calculation of these quantities require the solution of various field-induced **time-independent or time-dependent many-electron problems** (MEPs).

 The paper reviews briefly elements from our work on the solution of such MEPs, which has been developed and carried out within *state- and property-specific* frameworks. The discussion focuses on only a few items, such as, **1)** An outline on the calculation of state-specific wavefunctions including electron correlations and interchannel couplings. **2)** An argument/recipe regarding the systematic selection and calculation of the most important zero-order and correlation symmetry-adapted configurations for the calculation of one-photon transition amplitudes and the corresponding transition probabilities. **3)** An outline of the *state-specific expansion approach*, introduced in 1993-94, to the nonperturbative solution of the *many-electron* time-dependent Schrödinger equation with an initial state whose electronic structure may be labeled, in zero-order, by a single or a multiconfigurational wavefunction.

The material is presented as a commentary with explanations. The formal details and the numerical results can be found in the cited publications.

## 1. Introduction

Since the genesis of Quantum Mechanics, the quest for the understanding of properties and phenomena caused by the interaction of radiation with atoms and molecules has been on the frontier of basic research.

A very important component of this research is the pursuit of sound theories and of corresponding generally applicable computational methods that can solve reliably a large variety of field-induced *time-independent or time-dependent many-electron problems* (MEPs). This endeavor implies, among other things, the possibility of treating systematically the effects of **electron correlations** in N-electron systems on the observables.

The present paper reviews briefly aspects of our work in this area, where Hermitian as well as nonHermitian formalisms and methods of computation have been developed in the framework of the **state- and property-specific** (SPS) approach to the analysis and calculation of properties and phenomena in atomic and molecular physics, e.g., [1-5].

The results of these SPS approaches have demonstrated that it is possible to account quantitatively for the interplay between field-induced electron dynamics and the electronic structures of the states involved in each process of interest, while treating efficaciously the effects of the state-specific *self-consistent fields* and *electron correlations*. The electronic structures of the states involved may have N electrons, with N≥2, and their zero-order label may be single- or multi-configurational.

In other words, the theory and methodologies reviewed in detail in [1-5] allow the treatment of arbitrary electronic structures and are not restricted to the solution of problems that deal exclusively with the simple two-electron ground states of $He$, or of $H^-$ or of $H_2$, or of singlet ground states of two electrons outside a closed core. The need to go beyond such systems applies to both theory and experiment. For example, in one prototypical application which involved the ab initio solution of the many-electron time-dependent Schrödinger equation (METDSE), we underlined the significance of exploring field-induced time-dependent electron dynamics in N-electron systems with N > 2 and arbitrary electronic structures, as follows:

*"This is clearly a desideratum for experiment as well, if progress is to be characterized by new information on a variety of real systems. For example, only a small part of what one learns about control or about the effects of electron correlation from the experimental and/or theoretical studies of the interaction of the He $1s^2$ $^1S$ ground state with a strong electromagnetic pulse, can be transferred to other systems. In other words, just as in the case of the decades-old time-independent many-electron problem and the chasm between the many body problem for He and that for other larger atoms, especially in excited states, even a good calculation on the system "laser pulse— He $1s^2$ $^1S$ " cannot help very much in understanding quantitatively field-induced phenomena exhibited by arbitrary ground or excited states of other atoms across the Periodic Table".* (Page 1 of [6]).

The main focus in [6] was the prediction of the time-resolved formation of the channel-dependent resonances in the 13-electron Aluminum, after photo-ejection by femtosecond pulses of electrons from its inner subshells. The results were obtained by solving the METDSE via the **state-specific expansion approach** (SSEA) [2,4,5], the essence of which is described in section 3. In another application to a novel time-dependent MEP, the SSEA was combined with an appropriate theory and produced quantitative explanations of the measurements of *time delay in photoemission* of electrons from the 2$s$ and 2$p$ subshells of Neon [7] – see also [4].

Section 2 is a short commentary on the field-induced MEP.

Sections 3 and 4 refer to two different themes as examples of our SPS approaches to the solution of such MEPs. Section 3 is of direct relevance to modern time-dependent spectroscopy with ultrashort and strong radiation pulses. It has to do with the nonperturbative solution of the METDSE, which is achievable via the SSEA.

This endeavor, namely the nonperturbative solution of the METDSE, is currently an active area of theoretical and computational research, following the advances in the technology of producing intense and ultrashort radiation pulses with wavelengths ranging from the infrared to the soft X-ray regimes, and in experimental techniques that use them for purposes of time-resolved spectroscopy. A recent review of various methods aiming at solving the METDSE for such problems was published by Ishikawa and Sato [8]. Since, as they state in a note added in proof, the review did not include the SSEA, the present outline can be thought of as completing their discussion.

Section 4 goes back to the early stages of this research program in the 1970s, when there was strong experimental and theoretical interest in determining accurately one-photon transition rates and lifetimes of excited discrete states. It recalls the key features of the ***first-order theory of oscillator strengths*** (FOTOS) [9], whose implementation allows the identification and calculation of the effects of those state-specific one- and two- electron correlations which play the overwhelmingly dominant role in the transition process beyond the contribution of the zero-order wavefunctions.

Based on the theory of electronic structures for closed or open-shell states, and of the calculation of N-electron matrix elements, FOTOS justifies the systematic selection and calculation of the most important zero-order and correlation symmetry-adapted configurations for the calculation of one-photon transition amplitudes and the corresponding transition probabilities.

I note that, for reasons of economy, I leave out another area of investigations into field-induced MEPs. These have to do with the nonperturbative many-electron treatment of transition rates and of energy shifts that are caused by strong ac-type fields or by static fields. The solution to such MEPs can be pursued via the complex-eigenvalue, nonHermitian *many-electron, many-photon theory* (MEMPT), introduced in the late 1980s [1,5]. For a recent publication where the results of the MEMPT and those of the SSEA have been compared with measurements of the absolute cross-sections for the two-photon ionization of helium by free-electron laser pulses in the extreme ultraviolet , the reader is referred to [10].

## 2. Concerning the field-induced many-electron problem

*For every complex problem there is a solution that is clear, simple and wrong*

H. L. Mencken (1880-1956)

The fundamental equation which contains the information regarding the properties and phenomena which are observable when a pulse of radiation interacts with an atomic or molecular state, is the METDSE,

$$\boldsymbol{H}(t)\Psi(t) = i\hbar \frac{\partial \Psi(t)}{\partial t}, \qquad \boldsymbol{H}(t) = \mathbf{H} + \boldsymbol{V}(\omega,t) \qquad (1)$$

$\mathbf{H}$ is the N-electron field-free Hamiltonian, consisting of one- and two-electron nonrelativistic or relativistic operators. $\boldsymbol{V}(\omega,t)$ is the interaction one-electron operator. Apart from time and the photon frequency $\omega$, it contains information about the polarization and intensity of the field and the temporal characteristics of the pulse.

In our SSEA work on the solution of eq. 1, the electron-field coupling operator in $\boldsymbol{V}(\omega,t)$ has been used either in the electric dipole approximation or as the full electric operator of the *multipolar Hamiltonian* [11] constrained by dipole selection rules, [12] and its references.

Schrödinger's replacement $\Psi_i(q,t) = e^{-(i/\hbar)E_i t}\Psi_i(q)$ results in the Schrödinger equation for the N-electron discrete and energy-normalized scattering stationary states of $\mathbf{H}$,

$$\mathbf{H}\Psi_i(q) = E_i \Psi_i(q) \qquad (2)$$

*i* labels states of either the discrete or the continuous spectrum, and *q* stands for the coordinates of the N electrons.

Given the Hamiltonian in eq.1 and the two Schrödinger equations, the recipe of quantum mechanics for obtaining quantitative answers to all field-induced properties and phenomena with the Hamiltonian of eq. 1, is clear: For each MEP of interest, construct the theory and compute N-electron matrix elements with time-dependent or time-independent wavefunctions. The earliest such theories, appropriate for weak ac-fields, produce time-independent formulas from perturbation theory and use the stationary states of **H**, e.g., [13,14].

The task defined in the previous paragraph has challenged theoreticians since the early days of quantum mechanics. In this context, it should be recalled that most publications and books related to this subject *ignore* the serious and challenging issue of the proper solution of the MEP, either in time-dependent or in time-independent formulations. In other words, no N-electron wavefunctions are calculated and used. Instead, the contents of these publications are exhausted at levels of either pure formalism or of phenomenology. In such categories are the many discussions based on two- or three- or four-level models, where the N-electron matrix elements and energies are taken to be parameters.

In other types of investigations, the MEP is bypassed by conveniently replacing the true atomic Hamiltonian with the solvable *independent electron model* (IEM). In this way, no information as to the state-specific effects of *electron correlations* or *multichannel mixings* is obtained. Unless one has a good understanding from experience or from formal analysis as to the situations for which the IEM may be reasonably accurate and meaningful, predictions for arbitrary systems are unreliable.

## 2.1 The meaning of the term 'electron correlations'

It is useful to avoid misunderstandings by stating the meaning of the term 'electron correlations' which is used in this paper and is an integral part of the MEP. This term is not the same as 'electron interactions', whose meaning is expressed by the two-electron operators, nonrelativistic or relativistic.

The term *electron correlation* was coined by Wigner in his theory of the electron gas for metals in 1934 [15]. It was later picked up and first used in the literature of Quantum Chemistry in 1952 by Taylor and Parr [16], in their configuration-interaction study of the ground state of helium. It was introduced as the part of the exact solution of the ground state of eq. 2 which remains after the calculation of the self-consistent-field single-configuration wavefunction and energy. It should be noted that the MEP is widely known and discussed mostly in the context of the many-electron formalisms and, especially, of the countless computational applications having to do with the total energy and the properties of the *ground* states (or of a couple of low-lying discrete stationary states) of atoms and molecules. Given that for normal ground states the Hartree-Fock (HF) equations are solvable, this MEP has been essentially defined as the pursuit of the solution of $\mathbf{H}\Psi_g = E_g \Psi_g$ in the form $\Psi_g = \Phi_g^{HF} + X_g^{corr}$, $E_g = E_g^{HF} + E_g^{corr}$, which is the *electron correlation problem* with respect to the zero-order $\Phi_g^{HF}$, $E_g^{HF}$.

Even though the pursuit of the accurate determination of the total energy of ground states continues to characterize most of the publications on many-electron methods and algorithms, it should be kept in mind that for the purpose of understanding and computing properties and dynamics involving highly excited N-electron states, this is not necessarily rewarding or even feasible. Indeed, when it comes to complicated cases of electronic structures with a few open subshells or of excited states (including resonance and autoionizing states) where near-degeneracies are heavy, this definition of electron correlations has to be modified, since it is much more meaningful to compute the zero-order wavefunction as a self-consistent multiconfigurational expansion, based on the *state-specific 'Fermi-sea'* of the SPS theory [3,5]. Accordingly, the form in which the state-specific stationary wavefunctions, $\Psi_n$, are defined, computed and used is [3],

$$\Psi_n = a_0 \Psi_n^0 + \sum_{i=1}^{K} a_i X_n^i, \qquad a_0^2 + \sum_{i=1}^{K} a_i^2 = 1, \qquad K \to \infty \tag{3}$$

$$\equiv a_0 \Psi_n^0 + X_n^{corr} \tag{3a}$$

$\Psi_n^0$ is, in general, a self-consistently optimized multiconfigurational wavefunction containing the Fermi-sea configurations and $X_n^i$ are the remaining symmetry-adapted correlation configurations containing either one, or two, or three, etc., virtual excitations which are represented by analytic orbitals that are optimized variationally [3,9].

The SPS criterion for a good and physically meaningful calculation is that $a_0$ must be very close to unity. The number and nature of the correlation terms in the sum of eq. 3 then depend on the property of interest and on the level of accuracy required for meaningful comparison with experiment [3,5].

Here it must be added that when the problem requires the consideration of perturbed spectra and channel mixings, the theory which we have developed and applied for producing the perturbed wavefunctions becomes much more sophisticated, e.g., [17] and its references. Its structure is such that it can be applied (has been applied) not only to two-electron systems but to any type of electronic structures.

The main argument of the SPS theory [3, 5] is that in order to compute with accuracy the matrix element of an operator which represents a particular property other than the total energy, what matters most is the 'optimal matching' of the necessarily approximate wavefunction with the characteristics of the operator, and not whether the total energy has been obtained to a very good approximation.

Of course, for some operators the degree of convergence of the two calculations (first the total energy and then another property) is normally in harmony, especially for ground states. In general however, the aforementioned 'optimal matching' is not guaranteed by the criterion of the accuracy of the total energy, since non-negligible contributions to the value of the property may come from components of the wavefunction which do not contribute significantly to the total energy. These are configurations which have small coefficients, whilst their orbital radials are evidently not optimized with respect to the matrix element of interest. Yet, they are responsible for cumulative effects involving constructive or destructive contributions with quantitative consequences.

The preceding observations suggest that, in general, when it comes to the calculation of properties and phenomena, the treatment of the omnipresent MEP need not rely exclusively on huge calculations that first obtain a very accurate total energy, especially since these are most often impractical for problems involving field-induced nonstationary states. Instead, in addition to the basics of the energy criterion, it is rewarding to also focus on, and develop an understanding of the characteristics of the *N*-electron wavefunction(s) with respect to the *state-specific* self-consistent fields and the corresponding *state-specific electron correlations*. To this purpose, it is essential for theory not only to be formally correct and general, applicable to states of all kinds, but also to be implementable in terms of adequately optimized and numerically accurate *function spaces* [1-5].

As an example of the previous statements, I cite a case from our early (1970s) many-electron calculations of oscillator strengths of transitions where electron correlation is important. It belongs to the initial exploratory steps in the systematic treatment of various MEPs in terms of SPS-type approaches. The example has to do with the often discussed issue of the choice of the form of the electric-dipole transition operator in calculations of oscillator strengths. In a related analysis and ab initio calculations, we demonstrated how suitable choices of function spaces representing electron correlation can improve the results, even when the 'acceleration' formula, ($1/r^2$ dependence), is used, irrespective of whether the total energy for each of the two states of the transition (beyond the HF level) is accurate or not [18].

## 3. The nonperturbative solution of the METDSE via the state-specific expansion approach

When the intensity of the radiation pulse is strong with respect to the atomic state with which it interacts, theory must treat the MEP nonperturbatively. Furthermore, when the duration of the pulse is ultrashort, say in the range below 100 femtoseconds down to attoseconds, the physically and computationally significant issue of the interplay between dynamics and electronic structures and correlations must be resolved on the *time axis*.

The related critical question is how to solve nonperturbatively the METDSE using the many-electron **H** of eq. 1. Following the first implementation by Kulander in the late 1980s of the numerical space-time grid method to the problem of multiphoton ionization of noble gases at the one-electron level [19,20], in the early 1990s we examined how the METDSE could be solved for any electronic structure, including the effects of electron correlations and channel couplings. We concluded that a robust and general approach would be one that invokes the fundamental expansion principle of quantum mechanics and capitalizes on the knowledge and experience with the calculation of the stationary *state-specific* N-electron wavefunctions of the discrete and the continuous spectrum [2,4,5].

Specifically, the formal solution of eq. 1 can be written as, (using only one-channel for the continuum),

$$|\Psi(t)> = \sum_m a_m(t)|m> + \int_0^\infty b_\varepsilon(t)|\varepsilon> d\varepsilon \qquad (4)$$

where $|m>$ are the stationary states of the discrete spectrum and $|\varepsilon>$ are those of the continuous spectrum. The latter are energy-normalized according to $<\varepsilon|\varepsilon'> = \delta(\varepsilon - \varepsilon')$.

If the state-specific wavefunctions $|m>$ and $|\varepsilon>$ that are relevant to the problem of interest are available from separate calculations, then the objective is to calculate correctly the complex coefficients in eq. 4, and obtain a well-defined and usable N-electron $\Psi(t)$. The calculation of $\Psi(t)$ is achieved by solving the coupled equations which result from the substitution of eq. 4 into the METDSE. This is the essence of the *state-specific expansion approach* (SSEA) [2,4]. This endeavor requires the calculation of energies and of *bound-bound*, *bound-free* and *free-free* N-electron matrix elements using state-specific wavefunctions [2,12].

One of the advantages of constructing the expansion (4) in terms of state-specific wavefunctions is that, in the process of the solution, one can monitor and evaluate with transparency and economy the dependence of the evolution of $\Psi(t)$ on each $|m>$ and $|\varepsilon>$.

The solution of the METDSE according to the SSEA takes into account [2,4],

- The zero-order features of electronic structures of initial, intermediate and final states. These can be calculated at the HF or MCHF levels. (See eq. 3).
- The dominant electron correlations, for those states where this is necessary.
- The presence of perturbed or unperturbed Rydberg levels and multiply excited states.

- The contribution of the continuous spectrum of energy-normalized, channel-dependent scattering states, without or with resonance states.
- The interchannel coupling.

Since 1993-4, the SSEA has been implemented toward the successful solution of a variety of prototypical and demanding time-dependent MEPs having to do with electron dynamics [2,4-7]. Its foundations and computational methods are conceptually clear and mathematically rigorous.

**3.1 Molecules**

Although the scope of this paper does not include field-induced excitations of molecular vibrational levels, it is worth pointing out that the SSEA has also been applied to molecular dissociation and association problems, using either box-normalized basis sets [21] or energy-normalized ones [22] for the description of the dissociation continuum.

## 4. Understanding and calculating the effects of electron correlations on one-photon transitions

In spite of the advances on the frontiers of electron dynamics related to multiphoton ionization, the interest in determining one-photon discrete-discrete or discrete-continuum transition rates continues, especially when it comes to their computation between states with complex electronic structures labeled by open subshells. In fact, the growth of useful information regarding the many-electron effects on one-photon matrix elements involving such structures, can be helpful in the much more demanding MEPs with which one is faced when, say, aiming at the nonperturbative solution of the METDSE for one- or multi-photon absorption in such systems. A calculation of this type was reported in 2007 for the inner-electron time-dependent ionization of Aluminum [6]. However, going to heavier atoms may require the consideration of huge wavefunctions that include *relativistic correlations*.

In order to appreciate better the level of complexity of the field-induced MEPs when it comes to arbitrary atomic electronic structures, such as those with many electrons and with occupied open subshells with *p*, *d* or *f* orbitals, I refer to the work of Beck and coworkers on the calculation of properties taking into account *relativistic electron correlation*. For example, more than a decade ago, Beck [23] reported remarkable relativistic calculations of the electric dipole oscillator strengths for 264 transitions in Fe II, between the lowest 12 J = 9/2 levels of the state labeled by $(3d^6 4s + 3d^7)$ and 22 J = 9/2$^o$ levels of the $3d^6 4p$ state.

In such advanced calculations of transition probabilities [23], apart from having to handle efficiently the complexity due to relativistic electron couplings and electron correlations, it is crucial to have guidelines as to the choice of the zero-order and the correlation configurations which affect the transition probability the most, while keeping its calculation reliable. Otherwise, if the calculation follows the procedure of first solving eq. (2) for the total energy and wavefunction of the initial and the final states and then computing the absolute square of the transition matrix element, the task faces huge and, as it turns out, unnecessary difficulties. Instead, the work in [23] became feasible and cost-effective because the wavefunctions which were calculated and used contained overwhelmingly those correlation configurations that are dictated by FOTOS [9a,b]. In this section, I recall certain features of FOTOS, which offers an example of the SPS approaches to the quantitative understanding of field-induced processes. For longer discussions and applications, the reader is referred to [3,9,18].

The early proposals for both nonrelativistic and relativistic FOTOS procedures were published in 1975-78 [9a,b]. The first stages involved nonrelativistic theory. Later on, relativistic theory at the level of the Breit-Pauli Hamiltonian was implemented. Finally, with Beck's developments, the domain of relativistic correlation with Dirac spinors was reached. In the section on 'relativistic FOTOS' of [9b], we outlined procedures for three different possible types of relativistic formulations. For the most advanced one the instructions were written as follows: '*Define the fully relativistic Fermi-sea wavefunctions by using Fock-Dirac spinor configurations and including the Breit operator in the small configuration interaction. The relativistic first-order symmetries and correlations vectors are then derived by applying the relativistic selection rules for the transition of interest. The relativistic FOTOS wavefunctions now include Fermi-sea and virtual spinors whose optimization can be carried out variationally by projecting out the positron solutions. The fully* *relativistic transition operators and the experimental wavelength should be used*' (Page 178 of [9b]).

The concluding statement of the relativistic FOTOS section of [9b] reads: '*The efficient and correct evaluation of atomic and molecular transition probabilities within a rigorous relativistic correlation theory is certainly one of the most challenging new directions of Quantum Chemistry*' (Page 178 of [9b]). More than forty years later, the FOTOS prescription and results remain valid and relevant. Some of its characteristics are briefly reviewed below.

**4.1 First order theory of oscillator strengths (FOTOS)**

Whether in linear or non-linear responses to external radiation, the theory requires the calculation of N-electron matrix elements between pairs of states. Given that the radial characteristics of the coupling operator are different than those of **H**, the assertion can be made that although all electron correlations are needed to produce the exact solutions of eq. 2, this is not necessary when it comes to the calculation of reasonably accurate (i.e., meaningful quantitative comparison with experiment), coupling matrix elements needed for the calculation of a field-induced process. So the question that we raised in [9a,b] with respect to the calculation of one-photon oscillator strengths is:

**How can one go about identifying and computing the most important parts of the wavefunctions with respect to each N-electron transition strength of interest ?**

Such a possibility would not only facilitate the calculations but would also permit a general understanding as regards the contributions of electron correlations to field-induced transitions.

The development of FOTOS came as an answer to the previous question. In [9a,b] we proposed formal conditions and computational methods in terms of which one can approximate the exact transition strength, $S$, of the transition operator, $T$, with one which contains approximate wavefunctions which are much 'smaller' than the exact ones and, therefore, do not give the exact total energy of eq. 2. In spite of this, we expect to have,

$$S \equiv |<\Psi_i^{exact}|T|\Psi_f^{exact}>|^2 \approx |<\Psi_i^{FOTOS}|T|\Psi_f^{FOTOS}>|^2 \qquad (5)$$

The above relation implies that FOTOS provides the theoretical justification for determining the accurate values of the mixing coefficients for those zero-order ('Fermi-sea') and correlation configurations which contribute to S overwhelmingly. That a relationship such as (5) has meaning even for highly correlated wavefunctions can be realized by noting that a huge number of correlation configurations which contribute to the total energy of each state wavefunction have very small coefficients. When used in the $S$, the contribution of the corresponding matrix elements is negligible.

The analysis in [9a,b] is based on,

1) The choice of the self-consistent multiconfigurational Fermi-sea (F-S) zero-order wavefunction for each state. The notion of the F-S wavefunction for ground or excited states, its significance to the theory of electronic structure, and its background, are discussed in [3].

2) The formal use of first perturbation theory in order to explore the *forms* and the orbital symmetries of the perturbed wavefunctions for initial and final state wavefunctions. This was done by analyzing the **transition amplitude** of eq. 2 of [9a]:

$$<\Psi_i|T|\Psi_f> = <\Psi_i^0|T|\Psi_f^0> + <\Psi_i^0|T|\Psi_f^1> + <\Psi_f^0|T|\Psi_i^1> +$$

$$+ <\Psi_i^1|T|\Psi_f^1> \qquad (6)$$

The significant terms which are present in the forms of $\Psi_i^1$ and $\Psi_f^1$, are calculated variationally to all orders, for each state separately, using different function spaces.

I close by making three comments in connection with FOTOS.

*i)* The FOTOS wavefunctions are relatively 'small', since they do not include all the terms that contribute to the total energy. In cases of severe near-degeneracies, as it happens in relativistic spectra or in valence - Rydberg states mixed spectra, it is necessary to make an empirically determined small shift to the diagonal energies of the trial energy matrix so as to obtain the correct mixing coefficients of the dominant configurations. This type of diagonal –energy adjustment, which proved necessary in Beck's relativistic calculations of [23], was introduced in 1977 in the context of FOTOS for the easy correction of the main coefficients in heavily mixed wavefunctions [24]. It was later adopted and applied successfully to atomic spectra by Hibbert and coworkers, under the name '*fine tuning*' , e.g., [25].

*2)* Because of the separate optimization of the function spaces of the initial and the final wavefunctions, there is **nonorthonormality** (NON) between the initial and final orbital basis sets, at both the zero-order and the electron correlation levels. The FOTOS analysis showed that one of the consequences of this NON is that one- as well as two-electron excitations (or even multiple ones) with respect to the zero-order configurations may be allowed by orbital symmetry, even though $T$ is a one-electron operator. These excitations correspond to the virtual correlation space of the opposite state or to real excited states. This fact has computational as well physical consequences. For example, in conjunction with the multiconfigurational Fermi-sea zero-order wavefunction, it allows heuristic predictions of, say, satellite photoelectron peaks in addition to the main one [9a,b].

The systematic calculation of NON in the theory of transition probabilities in many-electron atoms was introduced in 1969 by Westhaus and Sinanoğlu for electric dipole transitions, using correlated wavefunctions that included the *nondynamical* correlations of first row atoms [26]. The methodology was extended in 1971 to electric quadrupole transitions by Nicolaides, Sinanoğlu and Westhaus [27]. The method was refined and optimized in the context of FOTOS (by processing symmetry-adapted configurations rather than determinants) in [9]. The first such demonstration to molecular transitions was published by Theodorakopoulos, Petsalakis and Nicolaides [28].

*3)*  As is well known, many observed features in spectra of discrete transitions or of photoionization, are attributed to excitations of two or more electrons from the initial state. The standard lore says that this type of electron dynamics is due *exclusively* to electron correlations. This assertion makes the problem look much harder than it actually is.

The main theoretical reason for this belief is that some formalisms, such as many-body perturbation theory, or the algorithm of the random phase approximation, or conventional configuration interaction, use a single basis set for initial and final states. Hence, the one-electron dipole operator cannot generate multiple excitations. However, as argued within the SPS theory, the electronic structures of excited states are best described in terms of their own function space and not in terms of the one that is optimized for, say, the ground state.

It is true that for any radiative transition electron correlation plays, by definition, a small or large role, both for the phenomenology and for the quantitative explanation. However, under certain conditions of orbital symmetry in the single configuration HF description, such transitions can be understood to a semi-quantitative level in terms of the '***Hartree-Fock transition theory***', without necessarily involving terms of electron correlations [3, 9a,b, 29]. Instead, the phenomenon results naturally as a consequence of the state-specific calculation of the HF wavefunctions for two different states. The corresponding transition matrix element is an N-electron integral. Therefore, because of NON between the two sets of HF orbitals, the presence of (N-1)-electron overlap integrals multiplied by the one-electron transition integral will give a non-zero result. This fact is valid only if the orbital symmetries in the overlap integral match completely.

For example, as a demonstration of this argument, in [29] I showed that a quantitatively reasonable two-electron photodetachment cross-section for the process $He^- \, 1s2s2p \,\, ^4P^o \rightarrow 1s\varepsilon p\varepsilon'p \,\, ^4P$ can be obtained with just a HF representation of the initial state. (Figure 1 of [29]). Of course, this does not negate the fact that when high accuracy is required, electron correlations must be included. In fact, this is what has been done for the same process near threshold using the MEMPT [30].

Even multiple (more than two) electron excitations can be described in zero-order without invoking electron correlations. For example, according to the HF transition theory, the electric dipole three-electron transition in Aluminum, $KL3s^2 3p \,\, ^2P^o \rightarrow KL4s^2 3d \,\, ^2D$, is 'easily' found to have the very small, (as expected), but *not* zero oscillator strength, when the state-specific HF equations are solved and the resulting HF wavefunctions are used for the calculation (Table 2 of [29]).

As a final example, I take the one from page 238 of [1], which discusses the term-dependent, one-photon three-electron transitions in Carbon for the multiplets $1s^2 2s^2 2p^2\ ^3P \rightarrow 1s^2 2s\ 3p^3\ ^3D^o, ^3P^o, ^3S^o$. The oscillator strengths for these transitions were calculated in terms of state-specific wavefunctions, without, (i.e., just the HF wavefunctions), and with electron correlations. The results show that the effect of electron correlations on the oscillator strengths is small, the values of the oscillator strengths being of the order of $10^{-4}$ [Page 238 of [1]). Hence, the description of the physical process is simple and transparent. On the contrary, if another type of formalism for electron dynamics were to be applied, the independent electron model would give zero for the oscillator strength and the calculation would have to go to high order (if possible) in order to pick up the correct three-electron excitation amplitude.

## 5. Conclusion

The implementation of *state-and property-specific* (SPS) many-electron approaches have produced reliable and transparent solutions to a variety of *time-independent* or *time-dependent* MEPs arising when many-electron atomic and molecular states interact with weak or strong radiation pulses.

The core argument and emphasis of the SPS theory have to do with the significance of using the proper *form* of the sought after solutions, and of understanding and computing in terms of separately optimized function spaces the electronic wavefunction of each state of the discrete and the continuous spectrum. For example, the METDSE is solved nonperturbatively according to the SSEA, which was discussed in section 3.

For atoms, the energy-normalized scattering orbitals are obtained numerically in the term-dependent HF or MCHF (N-1)-electron core. For both atoms and molecules, each bound N-electron wavefunction is obtained in the form of eq. 3, where the 'Fermi-sea' $\Psi_n^0$ contains the major electronic structure components and the remaining sum results from the analysis of correlations and interchannel-couplings in the discrete and the continuous spectrum, expected to affect the property of interest the most. An important example of such analysis is the use of FOTOS for the calculation of one-photon N-electron matrix elements and transition probabilities, outlined in section 4.

The discussion was brief, for reasons of economy. Details and many results can be found in the cited references.